\documentclass[12pt]{article}
\usepackage{graphicx}
\usepackage[cp1251]{inputenc}
\usepackage{rotating}
\begin{document}
 \noindent {\footnotesize\it Astronomy Reports, 2022, Vol. 66, No 4, pp. 269--277}
 \newcommand{\dif}{\textrm{d}}

 \noindent
 \begin{tabular}{llllllllllllllllllllllllllllllllllllllllllllll}
 & & & & & & & & & & & & & & & & & & & & & & & & & & & & & & & & & & & & & &\\\hline\hline
 \end{tabular}

 \vskip 0.5cm
  \centerline{\bf\large Galaxy Rotation Parameters from OB2 Stars with Proper Motions}
  \centerline{\bf\large and Parallaxes from the Gaia EDR3 Catalog}
 \bigskip
 \bigskip
  \centerline
 {
 V.V. Bobylev and A.T. Bajkova
 }
 \bigskip
 \centerline{\small \it
 Central (Pulkovo) Astronomical Observatory, Russian Academy of Sciences,}
 \centerline{\small \it Pulkovskoe shosse 65, St. Petersburg, 196140 Russia}
 \bigskip
 \bigskip
 \bigskip

 {
{\bf Abstract}---We have analyzed the kinematics of OB2 stars with proper motions and parallaxes selected by Xu et al. from the Gaia EDR3 catalog. The relative parallax errors for all the stars in this sample do not exceed 10\%. Based on a sample of 9750 stars, the group velocity components $(U,V,W)_\odot=(7.21,7.46,8.52)\pm(0.13,0.20,0.10)$~km/s were obtained and the parameters of the angular velocity of rotation of the Galaxy:
 $\Omega_0 =29.712\pm0.062$~km/s/kpc,
 $\Omega^{'}_0=-4.014\pm0.018$~km/s/kpc$^{2}$ and
 $\Omega^{''}_0=0.674\pm0.009$~km/s/kpc$^{3}$.
The circular velocity of rotation of the solar neighborhood around the center of the Galaxy is $V_0=240.7\pm3.0$~km/s for the assumed distance of the Sun to the galactic center $R_0=8.1\pm0.1$~kpc. It is shown that the influence of the systematic correction to the trigonometric parallaxes of the Gaia EDR3 catalog with the value $\Delta\pi=-0.040$~mas does not exceed the $\sim1\sigma$ level of the errors of the sought-for kinematic parameters of the model. Based on the proper motions of OB stars, the following variances of the residual
velocities were found: $(\sigma_1,\sigma_2,\sigma_3)=(11.79,9.66,7.21)\pm(0.06,0.05,0.04)$~km/s. It is shown that the first axis of this ellipsoid slightly deviates from the direction to the center of the Galaxy $L_1=12.4\pm0.1^\circ$, and the third axis is oriented almost exactly to the north pole of the Galaxy, $B_3=87.7\pm0.1^\circ.$
  }

\medskip DOI: 10.1134/S1063772922040011

 \section{INTRODUCTION}
Stars of spectral classes O and early B are very
young (few million years) massive (more than 10$M_\odot$)
stars of high luminosity. Due to these properties, they
hold great significance for the studies of the structure
and kinematics of the Galaxy at various scales.

OB stars are used to study the structure and kinematics
of the solar neighborhood, which harbors
young open clusters [1], OB associations [2--4], the
Gould Belt [5, 6], and the Local Arm [7].

There is a large number of known so-called runaway
stars. These are mainly OB stars that left their
parent cluster or association at high velocities [8--11].

Due to their high luminosity, OB stars are visible
from very far distances from the Sun. Spectrophotometric
distances are estimated from OB stars with relative
errors of 15--25\% [12--14]; until recently, those
served as the main source of distances to these stars.
Many O stars are surrounded by compact shells of ionized
hydrogen, or the so-called HII zones. The HII
zones and OB stars trace the large-scale structure of
the Galaxy well. For example, they are used to study
the curvature of the thin disk [15, 16] or the galactic
spiral pattern [7, 14, 16--19].

OB stars are used to determine the parameters of
galactic rotation [3, 20--31]. Often, since the radial
velocities of single OB stars are measured with large
errors, only their proper motions are analyzed.

As part of the Gaia space experiment [33], a version of the Gaia EDR3 catalog (Gaia Early Data Release 3 [34]) was published, in which the values of
trigonometric parallaxes and proper motions for about
1.5 billion stars were refined by approximately 30\% as compared with the previous version, Gaia DR2 [35]. Trigonometric parallaxes for about 500 million stars
were measured with errors less than 0.2 mas~\footnote [1]{mas is milliarcsecond}. For stars
with magnitudes $G<15^m$, random errors in the measurement
of proper motions lie in the range of 0.02--0.04 mas/year, and they greatly increase for fainter stars. In general, the proper motions of about half of
the stars in the catalog were measured with a relative
error of less than 10\%. There are no new radial velocity
measurements in the Gaia EDR3 catalog.

Xu et al. [19] presented a catalog of 5772 stars of spectral classes O--B2, in which the proper motions and trigonometric parallaxes of the stars were taken from the Gaia DR2 catalog. The kinematic analysis of these OB stars was performed by Bobylev and Bajkova
[32]. In [7], Xu et al. compiled a new, larger sample of OB stars with the proper motions and trigonometric parallaxes from the Gaia EDR3 catalog. The aim of the present study is to redefine the parameters of the rotation of the Galaxy using the latest data on stars of
spectral classes O and B from [7].

\section{METHODS}
\subsection{Galaxy Rotation Parameters}
From observations, we know three components of a star's velocity: radial velocity $V_r$ and two tangential velocity projections $V_l=4.74r\mu_l\cos b$ and $V_b=4.74r\mu_b$
oriented along the galactic longitude $l$ and latitude $b$, respectively, and expressed in km/s. The factor 4.74 is the dimension coefficient, and is the heliocentric distance of the star $r$ in kpc, which is calculated via parallax as $\pi$ as $r=1/\pi$. The proper motion
components and are expressed in mas/year.

To determine the parameters of the galactic rotation
curve, we use the equations obtained from the
Bottlinger formulas, in which the angular velocity $\Omega$ is
expanded in series up to terms of the second order of
smallness $r/R_0:$
\begin{equation}
 \begin{array}{lll}
 V_r=-U_\odot\cos b\cos l-V_\odot\cos b\sin l-W_\odot\sin b\\
 +R_0(R-R_0)\sin l\cos b\Omega^\prime_0
 +0.5R_0(R-R_0)^2\sin l\cos b\Omega^{\prime\prime}_0,
 \label{EQ-1}
 \end{array}
 \end{equation}
 \begin{equation}
 \begin{array}{lll}
 V_l= U_\odot\sin l-V_\odot\cos l-r\Omega_0\cos b\\
 +(R-R_0)(R_0\cos l-r\cos b)\Omega^\prime_0
 +0.5(R-R_0)^2(R_0\cos l-r\cos b)\Omega^{\prime\prime}_0,
 \label{EQ-2}
 \end{array}
 \end{equation}
 \begin{equation}
 \begin{array}{lll}
 V_b=U_\odot\cos l\sin b + V_\odot\sin l \sin b-W_\odot\cos b\\
 -R_0(R-R_0)\sin l\sin b\Omega^\prime_0
    -0.5R_0(R-R_0)^2\sin l\sin b\Omega^{\prime\prime}_0,
 \label{EQ-3}
 \end{array}
 \end{equation}
where $R$ is the distance from the star to the rotation axis of the Galaxy, 
$R^2=r^2\cos^2 b-2R_0 r\cos b\cos l+R^2_0.$ Velocities $(U,V,W)_\odot$ are the average
group velocity of the sample; they are taken with the opposite sign and
reflect the peculiar motion of the Sun. $\Omega_0$ is the angular
velocity of rotation of the Galaxy at solar distance $R_0$; parameters $\Omega^{\prime}_0$ and $\Omega^{\prime\prime}_0$ are the corresponding
derivatives of the angular velocity.

Knowing the $\Omega_0$ and $R_0$ values, we can calculate the linear velocity of rotation of the Galaxy at a near solar distance, $V_0=R_0\Omega_0$. In this study, the value $R_0$ is
taken equal to $8.1\pm0.1$~kpc according to the review by Bobylev and Bajkova [36], in which it was derived as a weighted average from a large number of modern individual estimates.

\subsection{Residual Velocity Ellipsoid}
The variance of the residual velocities of the stars is
estimated using the following known method [32].
We consider six moments of the second order
$a,b,c, f,e,d:$
\begin{equation}
 \begin{array}{lll}
 a=\langle U^2\rangle-\langle U^2_\odot\rangle,\qquad\quad
 b=\langle V^2\rangle-\langle V^2_\odot\rangle,\qquad\quad
 c=\langle W^2\rangle-\langle W^2_\odot\rangle,\\
 f=\langle VW\rangle-\langle V_\odot W_\odot\rangle,\quad
 e=\langle WU\rangle-\langle W_\odot U_\odot\rangle,\quad
 d=\langle UV\rangle-\langle U_\odot V_\odot\rangle,
 \label{moments}
 \end{array}
 \end{equation}
which are the coefficients of the surface equation
 \begin{equation}
 ax^2+by^2+cz^2+2fyz+2ezx+2dxy=1,
 \end{equation}
as well as the components of the symmetric tensor of
moments of residual velocities
 \begin{equation}
 \left(\matrix {
  a& d & e\cr
  d& b & f\cr
  e& f & c\cr }\right).
 \label{ff-5}
 \end{equation}
In this paper, the attention is focused on the analysis
of the proper motions of OB stars; there are few radial
velocities in this sample, so to determine the elements
of the residual velocity tensor, we use the following
three equations:
\begin{equation}
 \begin{array}{lll}
 V^2_l= a\sin^2 l+b\cos^2 l\sin^2 l
 -2d\sin l\cos l,
  \label{EQsigm-1}
  \end{array} \end{equation}
\begin{equation} \begin{array}{lll}
 V^2_b= a\sin^2 b\cos^2 l+b\sin^2 b\sin^2 l+c\cos^2 b\\
 -2f\cos b\sin b\sin l
 -2e\cos b\sin b\cos l
 +2d\sin l\cos l\sin^2 b,
 \label{EQsigm-2}
 \end{array}
 \end{equation}
\begin{equation}
 \begin{array}{lll}
 V_lV_b= a\sin l\cos l\sin b +b\sin l\cos l\sin b\\
 +f\cos l\cos b-e\sin l\cos b
 +d(\sin^2 l\sin b-\cos^2\sin b),
 \label{EQsigm-3}
 \end{array}
 \end{equation}
which are solved by the least-squares method (LSM) with respect to six unknowns $a,b,c,f,e,d$. The eigenvalues of tensor (6) $\lambda_{1,2,3}$ are then found from the
solution of the secular equation
 \begin{equation}
 \left|\matrix
 {
a-\lambda&          d&        e\cr
       d & b-\lambda &        f\cr
       e &          f&c-\lambda\cr
 }
 \right|=0.
 \label{ff-7}
 \end{equation}
The eigenvalues of this equation are equal to the reciprocals
of the square semiaxes of the velocity moment
ellipsoid and, at the same time, the square semiaxes of
the residual velocity ellipsoid:
 \begin{equation}
 \begin{array}{lll}
 \lambda_1=\sigma^2_1, \lambda_2=\sigma^2_2, \lambda_3=\sigma^2_3,\qquad
 \lambda_1>\lambda_2>\lambda_3.
 \end{array}
 \end{equation}
Directions of the principal axes $L_{1,2,3}$ and $B_{1,2,3}$ of tensor
(10) are found from the relations
 \begin{equation}
 \tan L_{1,2,3}={{ef-(c-\lambda)d}\over {(b-\lambda)(c-\lambda)-f^2}},
 \label{ff-41}
 \end{equation}
 \begin{equation}
 \tan B_{1,2,3}={{(b-\lambda)e-df}\over{f^2-(b-\lambda)(c-\lambda)}}\cos L_{1,2,3}.
 \label{ff-42}
 \end{equation}

\begin{figure}[t]
{ \begin{center}
  \includegraphics[width=0.85\textwidth]{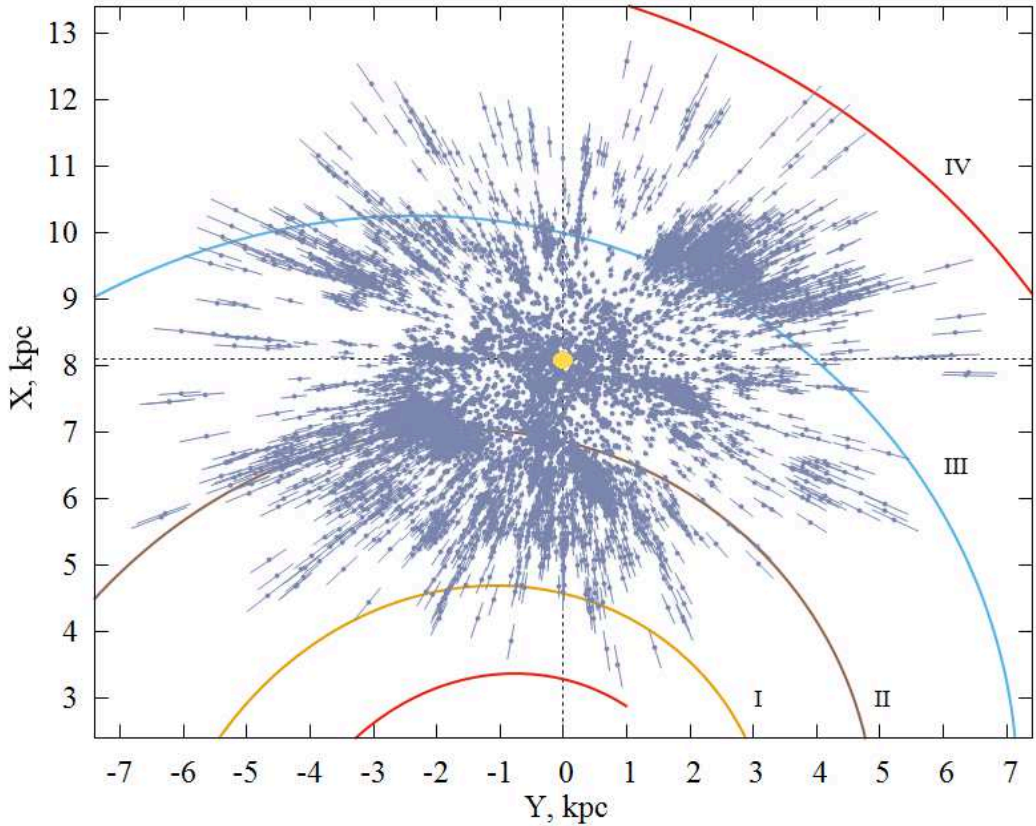}
  \caption{
Distribution of OB stars with relative parallax errors less than 7\% in projection onto the galactic plane $XY$; the position of the Sun is marked with a yellow circle, a four-armed spiral pattern with a twist angle $i=-13^\circ$ is shown according to [45].}
 \label{f-XY-7}
\end{center}}
\end{figure}
 \begin{table}[t] \caption[]{\small
Galaxy rotation parameters found from OB stars on the basis of equation (2) only,
$N_\star$ is the total number of the stars in the sample, $N_{eq}$ is the number of the equations.x
 }
  \begin{center}  \label{t:01}
  \small
  \begin{tabular}{|l|r|r|r|r|r|}\hline
  Parameters  & $\sigma_\pi/\pi<5\%$ & $\sigma_\pi/\pi<7\%$ & $\sigma_\pi/\pi<10\%$ \\\hline
     $N_\star$                   &          6861  &             8766  &          9750\\
      $N_{eq}$                   &          6764  &             8640  &          9610\\
     ${\overline z},$   pc       & $-14.5\pm1.0$  &     $-18.7\pm0.9$ & $-19.3\pm0.9$\\
     ${\overline r},$  kpc       &          1.89  &             2.12  &          2.27\\
                                 &&&\\
    $U_\odot,$    km/s           &   $ 6.80\pm0.19$ & $ 6.92\pm0.17$ & $  7.17\pm0.16$\\
    $V_\odot,$    km/s           &   $ 6.76\pm0.37$ & $ 7.43\pm0.29$ & $  7.37\pm0.24$\\

  $\Omega_0,$     km/s/kpc       & $29.633\pm0.084$ & $29.696\pm0.076$ & $29.700\pm0.076$\\
  $\Omega^{'}_0,$ km/s/kpc$^{2}$ & $-4.013\pm0.023$ & $-4.007\pm0.022$ & $-4.008\pm0.022$\\
 $\Omega^{''}_0,$ km/s/kpc$^{3}$ & $ 0.655\pm0.018$ & $ 0.670\pm0.011$ & $ 0.671\pm0.011$\\

   $\sigma_0,$    km/s           &           11.3  &            11.6  &           11.8\\

         $A,$    km/s/kpc        & $ 16.37\pm0.23$ &  $ 16.25\pm0.22$ &  $ 16.23\pm0.22$\\
         $B,$    km/s/kpc        & $-13.29\pm0.25$ &  $-13.38\pm0.24$ &  $-13.47\pm0.23$\\
         $V_0,$  km/s            & $ 240.3\pm3.1$  &  $ 240.0\pm3.0$  &  $ 240.6\pm3.0$ \\
  \hline
 \end{tabular}\end{center} \end{table}
 \begin{table}[t] \caption[]{\small
Galaxy rotation parameters found from OB stars using two equations of the form (2), (3), $N_\star$ is the total number of the stars in the sample, $N_{eq}$ is the number of the equations.
 }
  \begin{center}  \label{t:02}
  \small
  \begin{tabular}{|l|r|r|r|r|r|}\hline
  Parameters  & $\sigma_\pi/\pi<5\%$ & $\sigma_\pi/\pi<7\%$ & $\sigma_\pi/\pi<10\%$ \\\hline
     $N_\star$                   &          6861  &             8766  &          9750\\
      $N_{eq}$                   &         13513  &            17263  &         19202\\
                                 &&&\\
    $U_\odot,$    km/s           &   $ 6.90\pm0.15$ & $ 7.00\pm0.14$ & $  7.21\pm0.13$\\
    $V_\odot,$    km/s           &   $ 7.00\pm0.30$ & $ 7.57\pm0.24$ & $  7.46\pm0.20$\\
    $W_\odot,$    km/s           &   $ 8.27\pm0.11$ & $ 8.53\pm0.10$ & $  8.52\pm0.10$\\

  $\Omega_0,$     km/s/kpc       & $29.650\pm0.069$ & $29.704\pm0.062$ & $29.712\pm0.062$\\
  $\Omega^{'}_0,$ km/s/kpc$^{2}$ & $-4.022\pm0.019$ & $-4.013\pm0.018$ & $-4.014\pm0.018$\\
 $\Omega^{''}_0,$ km/s/kpc$^{3}$ & $ 0.666\pm0.015$ & $ 0.674\pm0.009$ & $ 0.674\pm0.009$\\

   $\sigma_0,$    km/s           &            9.2  &            9.4  &           9.6\\

         $A,$    km/s/kpc        & $ 16.39\pm0.23$ & $ 16.29\pm0.22$ & $ 16.26\pm0.21$\\
         $B,$    km/s/kpc        & $-13.22\pm0.24$ & $-13.36\pm0.23$ & $-13.45\pm0.22$\\
         $V_0,$  km/s            & $ 239.9\pm3.0 $ & $ 240.2\pm3.0$  & $ 240.7\pm3.0$ \\
  \hline
 \end{tabular}\end{center} \end{table}

\section{DATA}
In this paper, we used a sample of OB stars from the compilation by Xu et al. [7], for which the proper motions and trigonometric parallaxes were taken from the Gaia EDR3 catalog. For this purpose, 9750 stars of spectral classes O to B2, spectroscopically confirmed
by Skiff [37], were identified in [7] with the Gaia EDR3 catalog. The authors of [7] selected stars with relative errors of trigonometric parallaxes less than 10\%, while stars with pc were excluded from the sample.

The parallaxes of the Gaia EDR3 catalog apparently retained a small systematic shift with respect to the inertial coordinate system [38--43]. This shift was
first revealed in the Gaia DR2 parallaxes with a value
$\Delta\pi=-0.029$~mas [44], and later it was confirmed
from the analysis of various highly accurate data. This correction should be added to the measured parallaxes, so the true distances to the stars should be smaller. The $\Delta\pi$ correction to the parallaxes of the Gaia EDR3 catalog ranges from $-0.015$ [41] to $-0.039$~mas [40]. The value of the correction greatly
depends on the stellar magnitude, and it cannot be completely eliminated by simple methods.

Xu et al. [7] studied the effect of the correction $\Delta\pi=-0.017$~mas on the characteristics of the spiral pattern. The authors concluded that this systematic
correction does not significantly affect the character
of the spatial distribution of the OB stars under study.
In this paper, we aim to verify the influence of the correction
on the sought-for kinematic parameters of OB stars.

Figure 1 shows the distribution of OB stars with relative parallax errors below 7\% in projection onto the galactic plane $XY$. The $X$ axis of the coordinate system
is oriented from the center of the Galaxy to the
Sun, and the direction of the $Y$ axis coincides with the
direction of rotation of the Galaxy. A four-arm spiral
pattern with a twist angle $i=-13^\circ$ [45] is constructed
for kpc; the following segments of the spiral arms are numbered with Roman numerals: I---Scutum, II---Carina–Sagittarius, III---Perseus, and IV---Outer Arm.

\section{RESULTS}
The system of conditional equations of the form (1)--(3) is solved by the least-squares method with weights $w_{r,l,b}=S_0/\sqrt {S_0^2+\sigma^2_{V_{r,l,b}}}$, where $S_0$ is the ``cosmic'' variance, and $\sigma_{V_r},\sigma_{V_l}, \sigma_{V_b}$ are the error variances of the corresponding observed velocities. The value $S_0$ is comparable to the root-mean-square residual $\sigma_0$ (unit weight error) when solving conditional equations
of the form (1)--(3). We adopted $S_0=10$~km/s. The system of equations was solved in several iterations using the $3\sigma$ criterion to exclude open clusters with
large residuals.

The first way is to find a solution using a single conditional equation (2). The galactic rotation parameters found for three samples of OB stars with different
levels of parallax errors are listed in Table 1. The average
value of the ${\overline z}$ coordinate is given for each sample
(it reflects the ``elevation effect'', i.e., the altitude of
the Sun above the galactic plane). The obtained estimates
of ${\overline z}$ are in very good agreement, for example,
with the value ${\overline z}=-23\pm3$~pc found from the analysis
of open clusters with the data from the Gaia DR2 catalog in [46].

The values of the Oort constants $A=0.5\Omega'_0 R_0$ and $B=A-\Omega_0,$
calculated from the obtained $\Omega_0$ and $\Omega'_0$
values are given at the bottom of the table. The linear
velocity of rotation of the Galaxy at a near-solar distance,
$V_0=R_0\Omega_0$, is also given for the adopted value $R_0=8.1\pm0.1$~kpc.

To verify the influence of the systematic correction
to the parallaxes of the Gaia EDR3 catalog stars on
the kinematic parameters of OB stars, we use two
values, 0.020 and 0.040 mas. Using this method, for
the entire sample of 9750 stars with corrected parallaxes
$\pi=\pi+0.020$~mas, the velocity components
$(U,V)_\odot=(6.96,7.74)\pm(0.16,0.24)$~km/s and the following
parameters of the angular velocity of the galactic
rotation are found:
 \begin{equation}
 \label{solution-I}
 \begin{array}{lll}
      \Omega_0 =29.469\pm0.076~\hbox{km/s/kpc},\\
  \Omega^{'}_0 =-3.965\pm0.021~\hbox{km/s/kpc$^{2}$},\\
 \Omega^{''}_0 =~0.663\pm0.013~\hbox{km/s/kpc$^{3}$}.
 \end{array}
 \end{equation}
In this solution, the unit weight error $\sigma_0=11.2$~km/s.
The linear velocity of rotation of the Galaxy at a near-solar
distance $V_0=238.7\pm3.0$~km/s, and the Oort
constants $A=16.06\pm0.22$~km/s/kpc and $B=-13.41\pm0.23$~km/s/kpc.

Repeating solution (14) with new parallax values $\pi=\pi+0.040$~mas gives $(U,V)_\odot=(6.79,7.99)\pm(0.15,0.24)$~km/s and the parameters of the angular
velocity of the galactic rotation:
 \begin{equation}
 \label{solution-Ia}
 \begin{array}{lll}
      \Omega_0 =29.305\pm0.077~\hbox{km/s/kpc},\\
  \Omega^{'}_0 =-3.933\pm0.021~\hbox{km/s/kpc$^{2}$},\\
 \Omega^{''}_0 =~0.653\pm0.015~\hbox{km/s/kpc$^{3}$}.
 \end{array}
 \end{equation}
In this solution, the unit weight error $\sigma_0=10.7$~km/s. The linear velocity of rotation of the Galaxy at a near-solar distance $V_0=237.4\pm3.0$~km/s, and the Oort
constants $A=15.93\pm0.21$~km/s/kpc and $B=-13.38\pm0.23$~km/s/kpc. The values of parameters (14) and (15) should first be compared with the values from
the last column of Table~1, since they were found using the same stars.

The second way is to jointly solve the system of
conditional equations of the form (2)--(3). The Galactic
rotation parameters found by this method for three
samples of OB stars are given in Table~2. Using this
method, for the entire sample of OB stars with experimental
parallax correction $\pi=\pi+0.020$~mas, the following parameters were found: $(U,V,W)_\odot=(6.98,7.81,8.14)\pm(0.13,0.20,0.09)$~km/s and
 \begin{equation}
 \label{solution-II}
 \begin{array}{lll}
      \Omega_0 =29.461\pm0.062~\hbox{km/s/kpc},\\
  \Omega^{'}_0 =-3.969\pm0.018~\hbox{km/s/kpc$^{2}$},\\
 \Omega^{''}_0 =~0.665\pm0.011~\hbox{km/s/kpc$^{3}$}.
 \end{array}
 \end{equation}
In this solution, the unit weight error $\sigma_0=9.1$~km/s. The linear velocity of rotation of the Galaxy at a near-solar distance $V_0=238.6\pm3.0$~km/s, and the Oort
constants $A=16.07\pm0.21$~km/s/kpc and $B=-13.39\pm0.22$~km/s/kpc. Values (16) should be compared with those given in the last column of Table 2.

The tables show the $N_{eq}$ value, which indicates the actual number of stars in the search for a solution after filtering by the $3\sigma$ criterion. In Table~1, the number of
stars rejected by this criterion is indicated by a simple
difference $N_\star-N_{eg}$. The number of discarded stars here is very small (less than 1\%). In the second case, the number of discarded stars is indicated by the difference $2N_\star-N_{eg}$, and this number is larger, but also not critical.

Xu et al. [7] formed a sample of 5772 O-B2 stars with kinematic parameters from the Gaia DR2 catalog. The radial velocities for more than 2500 of the stars were taken from the SIMBAD electronic database~\footnote [2]{http://simbad.u-strasbg.fr/simbad/}.

We identified the samples of OB stars from [7, 19]
and found 1812 stars with radial velocities in the new
sample. The radial velocities of OB stars in the catalog
of Xu et al. [7] are given relative to the local standard
of rest, so we convert them back to heliocentric velocities
with the known parameters of the standard
motion of the Sun $(U,V,W)_\odot=(10.3,15.3,7.7)$~km/s.

The interest in these stars is associated primarily with the fact that they can be used to plot the rotation curve of the Galaxy. To do this, we calculate the spatial
velocities $U,V,W$, and then another two velocities:
$V_R$, directed radially from the galactic center, and the
velocity orthogonal to it $V_{circ}$ in the direction of rotation
of the Galaxy based on the following relations:
 \begin{equation}
 \begin{array}{lll}
  V_{circ}= U\sin \theta+(V_0+V)\cos \theta, \\
       V_R=-U\cos \theta+(V_0+V)\sin \theta,
 \label{VRVT}
 \end{array}
 \end{equation}
where the position angle $\theta$ satisfies the relation
$\tan\theta=y/(R_0-x)$, and $x,y,z$ are the rectangular
heliocentric coordinates of the star (velocities
$U,V,W$ are oriented along the corresponding axes $x,y,z$).

It should be noted that in a sample of 1812 OB stars
with radial velocities, the errors in determining the
radial velocities are not given for more than half of the
stars; for a significant part of the stars, the errors in
determining the radial velocities exceed 10~km/s.

The availability of radial velocities allows us to
search for a joint solution of a system of three conditional
equations of the form (1)--(3). The following
parameters were found using this method for OB stars
with radial velocities and proper motions:
$(U,V,W)_\odot=(7.17,10.03,8.15)\pm(0.30,0.35,0.29)$~km/s and
 \begin{equation}
 \label{solution-RV}
 \begin{array}{lll}
      \Omega_0 =29.22\pm0.19~\hbox{km/s/kpc},\\
  \Omega^{'}_0 =-3.885\pm0.042~\hbox{km/s/kpc$^{2}$},\\
 \Omega^{''}_0 =~0.685\pm0.031~\hbox{km/s/kpc$^{3}$},
 \end{array}
 \end{equation}
where the unit weight error $\sigma_0=12.2$~km/s, and the
linear rotation velocity of the Galaxy at a near-solar
distance $V_0=236.7\pm3.3$~km/s. After discarding stars
with large radial velocity errors (more than 20~km/s),
as well as using the $3\sigma$ criterion, 1726 OB stars
remained, for which solution (18) was found and
Fig. 2 was plotted.

\begin{figure}[t]
{ \begin{center}
  \includegraphics[width=0.7\textwidth]{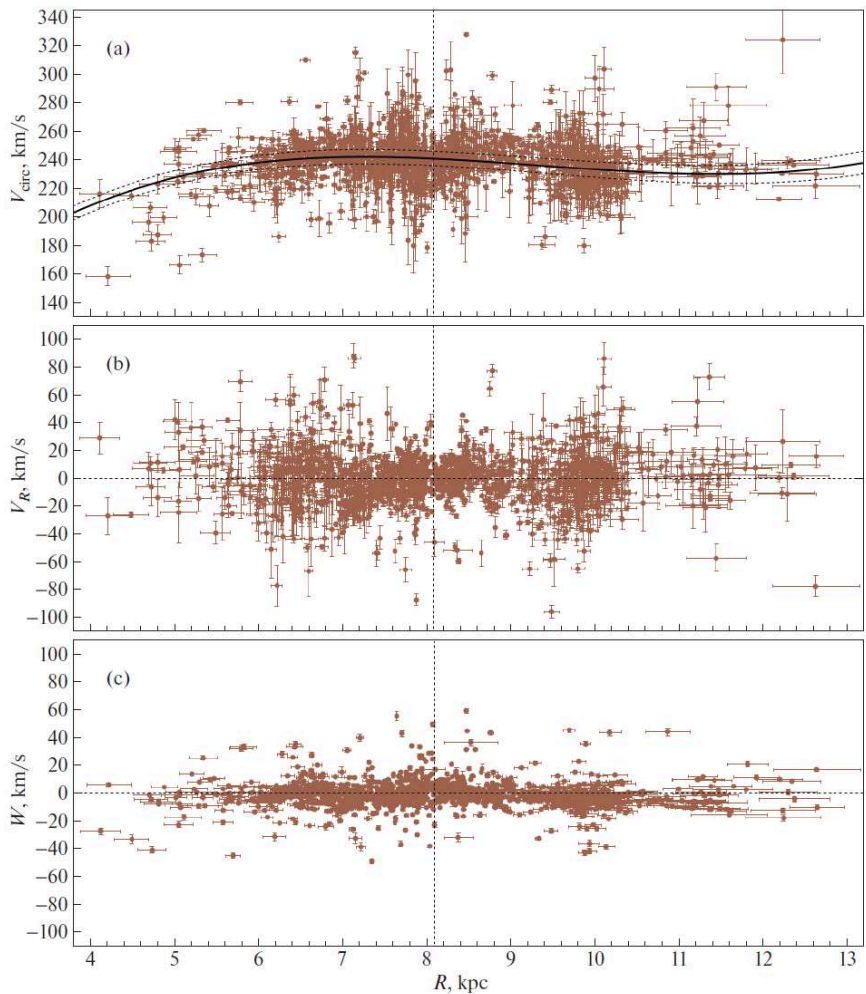}
  \caption{
Top panel (a): circular velocities of rotation of OB stars $V_{circ}$ as a function of distance $R$; the rotation curve with the boundaries of the confidence region corresponding to the $1\sigma$ level is given. Central panel (b): radial velocities $V_R$. Bottom panel (c): vertical velocities of OB stars $W$ as a function of distance $R$; the vertical line marks the position of the Sun.}
 \label{f-RV}
\end{center}}
\end{figure}

Figure 2 shows circular velocities of rotation $V_{circ},$
radial velocities $V_R$, and vertical velocities $W$ of 1726
OB stars depending on the distance $R$. The rotation
curve that we consider to be the best was found from
proper motions only (the last column of Table~2). As
can be seen from Fig. 2a, the rotation curve has a very
narrow confidence region.

Both in Fig. 2a and 2b, one can easily trace the wave-like behavior of the velocities, which is associated with the influence of the galactic spiral density
wave. Bobylev and Bajkova [32] performed a kinematic
Fourier analysis of more than 2000 OB stars
from the list [7], where Fig. 4 with periodic curves
describing the effect of the spiral density wave was plotted. Bobylev and Bajkova [32] found that the amplitudes of the tangential and radial perturbation velocities are $f_\theta=4.4\pm1.4$~km/s and $f_R=5.1\pm1.2$~km/s, respectively.

Based on the proper motions of 9720 OB2 stars, as
a result of the LSM solution of the system of conditional
equations of the form (7)--(9), the following
variances of the residual velocities were found:
 \begin{equation}
 \begin{array}{lll}
  \sigma_1=11.79\pm0.06~\hbox{km/s}, \\
  \sigma_2=~9.66\pm0.05~\hbox{km/s},\\
  \sigma_3=~7.21\pm0.04~\hbox{km/s},
 \label{rez-155}
 \end{array}
 \end{equation}
as well as the orientation parameters of this ellipsoid:
 \begin{equation}
  \matrix {
  L_1=~12.4\pm0.1^\circ,&B_1=+0.5\pm0.1^\circ,\cr
  L_2=102.4\pm0.1^\circ,&B_2=+2.3\pm0.1^\circ,\cr
  L_3=271.2\pm0.1^\circ,&B_3=87.7\pm0.1^\circ.\cr
    }
 \label{rez-255}
 \end{equation}
We can conclude that these are indeed very young
stars, which are characterized by a small variance of
residual velocities. It should be noted that the average
value $(\sigma_1+\sigma_2+\sigma_3)/3=9.55$~km/s, which characterizes
the variance of the average spatial velocity, is close
to the unit weight error values $\sigma_0$, which are listed in
Table~2, and to our chosen value of the ``cosmic'' variance
$S_0=10$~km/s.

\section{DISCUSSION}
At present, it is believed that the most reliable components of the peculiar velocity of the Sun relative to the local standard of rest, $(U,V,W)_\odot=(11.1,12.2,7.3)\pm(0.7,0.5,0.4)$~km/s, are determined
by Sch\"onrich et al. [47]. The $U_\odot$ and $V_\odot$ velocity values
found in this paper for various samples of OB stars differ greatly from those found in [47]. As shown in [48], there is an influence of the galactic spiral density wave,
and the $U_\odot$ and $V_\odot$ velocities greatly depend on the
phase of the Sun in the density wave. As can be seen
from Fig.~1, OB stars are strongly concentrated toward
segments of the spiral arms, so the kinematics of these
stars has to be influenced by the spiral density wave.
Although we do not attach much importance to solution (18) in terms of estimating the rotation parameters, the $V_\odot$ velocity value here is closer to the one found by Sch\"onrich et al. [47].

An important parameter is the value of the linear velocity $V_0$. It is known that such objects of the thin disk of the Galaxy as hydrogen clouds, maser sources
in active star-formation regions, OB stars, young open
clusters, the youngest Cepheids, etc., rotate the fastest.

In [31], the estimate $V_0=231\pm5$~km/s for the adopted value $R_0=8.0\pm0.15$~kpc was obtained from the analysis of 495 OB stars from the Gaia DR2 catalog.
Mr\'oz et al. [49] obtained the estimate $V_0=233.6\pm2.8$~km/s for adopted $R_0=8.122\pm0.031$~kpc from the analysis of about 770 classical Cepheids. In
[50], the velocity $V_0=232.5\pm0.9$~km/s for adopted $R_0=8.122\pm0.031$~kpc was found with very high accuracy based on the sample of about 3500 classical Cepheids. In [51], $V_0=240\pm3$~km/s was found for the calculated value $R_0=8.27\pm0.10$~kpc from the
analysis of 800 Cepheids.

Rastorguev et al. [52] used data on 130 galactic masers with measured trigonometric parallaxes to find the components of the solar velocity
$(U_\odot,V_\odot)=(11.40,17.23)\pm(1.33,1.09)$~km/s and the following
values of the parameters of the rotation curve of the
Galaxy:  
 $\Omega_0=28.93\pm0.53$~km/s/kpc,
 $\Omega^{'}_0=-3.96\pm0.07$~km/s/kpc$^{2}$ and
 $\Omega^{''}_0=0.87\pm0.03$~km/s/kpc$^{3},$ $V_0=243\pm10$~km/s for the found value $R_0=8.40\pm0.12$~kpc.

Reid et al. [53], using a sample of 147 masers, found the following values of the two most important kinematic parameters: 
$R_0=8.15\pm0.15$~kpc and $\Omega_\odot=30.32\pm0.27$~km/s/kpc, where
$\Omega_\odot=\Omega_0+V_\odot/R.$
The velocity value $V_\odot=12.2$~km/s was taken from [47]. These authors used a method based on the series expansion of the linear velocity of rotation of the Galaxy.

Based on the proper motions of approximately
6000 OB stars from the list [19] with proper motions and
parallaxes from the Gaia DR2 catalog, the authors of
[32] found $(U_\odot,V_\odot)=(6.53,7.27)\pm(0.24,0.31)$~km/s,
 $\Omega_0 =29.70\pm0.11$~km/s/kpc,
 $\Omega^{'}_0=-4.035\pm0.031$~km/s/kpc$^{2}$ and
 $\Omega^{''}_0=0.620\pm0.014$~km/s/kpc$^{3}$, where
$V_0=238\pm5$~km/s for the adopted $R_0=8.0\pm0.15$~kpc.
It should be noted that these values
must be compared with parameters~(16), which are
obtained on the basis of a completely identical
approach. This comparison shows that the errors in determining the kinematic parameters~(16) are approximately 1.5 times smaller.

From 788 Cepheids from the list of Mr\'oz et al. [49]
with proper motions and radial velocities from the
Gaia DR2 catalog, the authors of [51] found
$(U_\odot,V_\odot,W_\odot)=(10.1,13.6,7.0)\pm(0.5,0.6,0.4)$~km/s, as
well as 
      $\Omega_0=29.05\pm0.15$~km/s/kpc,
   $\Omega^{'}_0=-3.789\pm0.045$~km/s/kpc$^{2}$,
  $\Omega^{''}_0=0.722\pm0.027$~km/s/kpc$^{3}$, at calculated $R_0=8.27\pm0.10$~kpc.

Thus, we can conclude that the parameters of the angular velocity of rotation of the Galaxy $\Omega_0,$ $\Omega^{'}_0$ and $\Omega^{''}_0$
found in this paper for OB stars are in good agreement
with the estimates of other authors and are determined
with high accuracy in our case.

There is interest [54--57] in the values of the Oort constants $A$ and $B$. These constants characterize the shape of the Galactic rotation curve in a small neighborhood of the Sun. According to our definitions, the sum $A+B=-\partial V_{circ}/\partial R$ indicates that the linear velocity of the galactic rotation $V_{circ}$ decreases in the solar neighborhood (a slight deflection of the rotation curve in the region $R=R_0$ in Fig.~2a), which is in
agreement with modern estimates of the character of the rotation of the Galaxy.

For example, Bovy [56], from the analysis of the proper motions and parallaxes of a local sample of 304267 main sequence stars in the Gaia DR1 catalog~[58], found $A=15.3\pm0.5$~km/s/kpc and $B=-11.9\pm0.4$~km/s/kpc, on the basis of which he obtained an estimate of the angular velocity of the rotation of the Galaxy $\Omega_0=27.1\pm0.5$~km/s/kpc and velocity $V_0=219\pm4$~km/s.

Based on a large sample of Gaia DR2 stars located in the Sun's neighborhood with a radius of 500~pc, the following estimates were obtained in [57]: 
$A=15.1\pm0.1$~km/s/kpc, $B=-13.4\pm0.1$~km/s/kpc and $\Omega_0=28.5\pm0.1$~km/s/kpc.

\section{CONCLUSIONS}
The kinematics of the Galaxy was studied using a sample of OB2 stars from the paper by Xu et al. [7] with proper motions and trigonometric parallaxes
from the Gaia EDR3 catalog. These very young stars
are located no higher than 300 pc above the galactic
plane and no farther than 5--6~kpc from the Sun (on
average, at a distance of approximately 2 kpc).

Two approaches to solving kinematic equations
were tested: (a) using only the component $V_l$ and (b)
using two components, $V_l$ and $V_b.$ It was shown that in
comparison with the first method, the second method
has a slight advantage in the possibility of estimating
velocity $W_\odot,$ as well as in reducing the level of errors of
the determined parameters.

It was shown that the influence of the systematic correction to the trigonometric parallaxes of the Gaia EDR3 catalog with the value $\Delta\pi=-0.040$~mas does
not exceed the level (approximately 1$\sigma$) of errors of the
sought-for kinematic parameters of the model. The
actual effect of the correction is that the values of such
parameters as $\Omega_0,$ $\Omega^{'}_0,$ $\Omega^{''}_0,$ and $V_0$ and become smaller (in
absolute value). The beneficial effect is a significant reduction of the unit weight error $\sigma_0$ in the search for the LSM solution of kinematic equations.

The kinematic equations were solved using three
constraints on the stellar parallax errors $\sigma_\pi/\pi:$ 10\%, 7\%, and 5\%. We concluded that there was almost no dependence of the determined kinematic parameters
on the level of parallax errors.

From the sample of 9750 OB stars, without introducing a correction to their parallaxes, we found the group velocity components
$(U,V,W)_\odot=(7.21,7.46,8.52)\pm(0.13,0.20,0.10)$~km/s and the
following parameters of the angular velocity of rotation
of the Galaxy:  $\Omega_0 =29.712\pm0.062$~km/s/kpc,
 $\Omega^{'}_0=-4.014\pm0.018$~km/s/kpc$^{2}$ and
 $\Omega^{''}_0=0.674\pm0.009$~km/s/kpc$^{3}$. 
The circular velocity of the rotation of the solar neighborhood around the center of the
Galaxy $V_0=240.7\pm3.0$~km/s for the adopted distance $R_0=8.1\pm0.1$~kpc. Based on 1726 OB stars with radial velocities and proper motions, the $V_{circ}$ and $V_R$ velocities were calculated, and a graph of the rotation curve was plotted with parameters found from proper motions only. This curve was shown to have a very narrow confidence region.

Based on the proper motions of 9720 OB stars, the following variances of residual velocities were determined:
$(\sigma_1,\sigma_2,\sigma_3)=(11.79,9.66,7.21)\pm(0.06,0.05,0.04)$~km/s. It was shown that the first axis of this ellipsoid slightly deviates from the direction to the center of the Galaxy, $L_1=12.4\pm0.1^\circ$, and the third axis is directed almost exactly to the north pole of the Galaxy, $B_3=87.7\pm0.1^\circ.$

\subsubsection*{ACKNOWLEDGMENTS}
The authors are grateful to the reviewer for the valuable
comments, which helped to improve the paper.

\subsubsection*{CONFLICT OF INTEREST}
The authors declare that they have no conflicts of interest.

\subsubsection*{REFERENCES}

 {\small
 \quad
~1. E. Piskunov, N. V. Kharchenko, S. R\"oser, E. Schilbach and R.-D. Scholz, Astron. Astrophys. 445, 545 (2006).

2. P. T. de Zeeuw, R. Hoogerwerf, and J. H. J. de Bruijne, Astron. J. 117, 354 (1999).

3. A. K. Dambis, A. M. Mel’nik, and A. S. Rastorguev, Astron. Lett. 27, 58 (2001).

4. M. Mel’nik and A. K. Dambis, Mon. Not. R. Astron. Soc. 472, 3887 (2017).

5. J. A. Frogel and R. Stothers, Astron. J. 82, 890 (1977).

6. J. Torra, D. Fern\'andez, and F. Figueras, Astron. Astrophys. 359, 82 (2000).

7. Y. Xu, L.G. Hou, S. Bian, et al., Astron. Astrophys. 645, L8 (2021).

8. A. Blaauw, Bull. Astron. Inst. Netherland 15, 265 (1961).

9. R. Hoogerwerf, J. H. J. de Bruijne, and P. T. de Zeeuw, Astrophys. J. 544, L133 (2000).

10. N. Tetzlaff, R. Neuh\"auser, and M. M. Hohle, Mon.
Not. R. Astron. Soc. 410, 190 (2011).

11. V. V. Bobylev and A. T. Bajkova, Astron. Lett. 47, 224 (2021).

12. M. Mohr-Smith, J. E. Drew, R. Napiwotzki, et al., Mon. Not. R. Astron. Soc. 465, 1807
(2017).

13. B.-Q. Chen, Y. Huang, L.-G. Hou, et al., Mon. Not. R. Astron. Soc. 487, 1400 (2019).

14. J. M. Shull and C. W. Danforth, Astrophys. J. 882, 180 (2019).

15. R. Drimmel, R. L. Smart, and M. G. Lattanzi, Astron. Astrophys. 354, 67 (2000).

16. D. Russeil, Astron. Astrophys. 397, 133 (2003).

17. Y. M. Georgelin and Y. P. Georgelin, Astron. Astrophys. 49, 57 (1976).

18. D. Fern\'andez, F. Figueras, and J. Torra, Astron. Astrophys. 372, 833 (2001).

19. Y. Xu, S. B. Bian, M. J. Reid, J. J. Li, et al., Astron. Astrophys. 616, L15 (2018).

20. J. Byl and M. W. Ovenden, Astrophys. J. 225, 496 (1978).

21. M. Miyamoto and Z. Zhu, Astron. J. 115, 1483 (1998).

22. M. Uemura, H. Ohashi, T. Hayakawa, et al., Publ. Astron. Soc. Jpn. 52, 143
(2000).

23. R. L. Branham, Astrophys. J. 570, 190 (2002).

24. R. L. Branham, Mon. Not. R. Astron. Soc. 370, 1393 (2006).

25. M. V. Zabolotskikh, A. S. Rastorguev, and A. K. Dambis, Astron. Lett. 28, 454 (2002).

26. M. E. Popova and A. V. Loktin, Astron. Lett. 31, 663 (2005).

27. Z. Zhu, Chin. J. Astron. Astrophys. 6, 363 (2006).

28. A. M. Mel'nik and A. K. Dambis, Mon. Not. R. Astron. Soc. 400, 518 (2009).

29. M. Melnik and A. K. Dambis, Astrophys. Space Sci. 365, 112 (2020).

30. G. A. Gontcharov, Astron. Lett. 38, 694 (2012).

31. V. V. Bobylev and A. T. Bajkova, Astron. Lett. 44, 676 (2018).

32. V. V. Bobylev and A. T. Bajkova, Astron. Lett. 45, 331 (2019).

33. T. Prusti, J. H. J. de Bruijne, A. G. A. Brown, et al., Astron. Astrophys. 595, A1 (2016).

34. A.G. A. Brown, A. Vallenari, T. Prusti, et al., Astron. Astrophys. 649, A1 (2021).

35. G. A. Brown, A. Vallenari, T. Prusti, et al., Astron. Astrophys. 616, A1 (2018).

36. V. V. Bobylev and A. T. Bajkova, Astron. Rep. 65, 498 (2021).

37. A. Skiff, VizieR Online Data Catalog, B/mk (2014).

38. L. Lindegren, U. Bastian, M. Biermann, et al., Astron. Astrophys. 616, A2 (2021).

39. F. Ren, X. Chen, H. Zhang, et al, Astrophys. J. Lett. 911, L20 (2021).

40. M. A. T. Groenewegen, Astron. Astrophys. 654, A20 (2021).

41. J. C. Zinn, Astron. J. 161, 214 (2021).

42. Y. Huang, H. Yuan, T. Beers, and H. Zhang, Astrophys. J. Lett. 910, L5 (2021).

43. J. Maiz Apell\'aniz; arXiv: 2110.01475 [astro-ph.IM] (2021).

44. L. Lindegren, J. Hernandez, A. Bombrun, et al., Astron. Astrophys. 616, A2 (2018).

45. V. V. Bobylev and A. T. Bajkova, Mon. Not. R. Astron. Soc. 437, 1549 (2014).

46. T. Cantat-Gaudin, F. Anders, A. Castro-Ginard, et al., Astron. Astrophys. 640, A1 (2020).

47. R. Sch\"onrich, J. J. Binney, and W. Dehnen, Mon. Not. R. Astron. Soc. 403, 1829 (2010).

48. V. V. Bobylev and A. T. Bajkova, Mon. Not. R. Astron.
Soc. 441, 142 (2014).

49. P. Mr\'oz, A. Udalski, D. M. Skowron, J. Skowron, et al., Astrophys. J. 870, L10 (2019).

50. Ablimit, G. Zhao, C. Flynn, and S. A. Bird, Astrophys. J. 895, L12 (2020).

51. V. V. Bobylev, A. T. Bajkova, A. S. Rastorguev, and M. V. Zabolotskikh, Mon. Not. R. Astron. Soc. 502, 4377 (2021).

52. A. S. Rastorguev, M. V. Zabolotskikh, A. K. Dambis, et al., Astrophys.
Bull. 72, 122 (2017).

53. M. J. Reid, K. M. Menten, A. Brunthaler, et al., Astrophys. J. 885, 131 (2019).

54. F. Mignard, Astron. Astrophys. 354, 522 (2000).

55. R. P. Olling and W. Dehnen, Astrophys. J. 599, 275 (2003).

56. Jo Bovy, Mon. Not. R. Astron. Soc. 468, L63 (2017).

57. Li, G. Zhao, and C. Yang, Astrophys. J. 872, 205 (2019).

58. A. G. A. Brown, A. Vallenari, T. Prusti, et al., Astron. Astrophys. 595, A2 (2016). 
 }
 \end{document}